\documentclass[10pt,conference]{IEEEtran}
\usepackage{cite}
\usepackage{amsmath,amsfonts}
\usepackage{algorithmic}
\usepackage{graphicx}
\usepackage{booktabs}
\usepackage[caption=false,font=footnotesize]{subfig}
\usepackage{textcomp}
\usepackage{xcolor}
\usepackage{url}
\def\BibTeX{{\rm B\kern-.05em{\sc i\kern-.025em b}\kern-.08em
    T\kern-.1667em\lower.7ex\hbox{E}\kern-.125emX}}

\begin{document}

\title{Coupling Planning with Episodic Memory in LLM Agents for Software Issue Resolution}

\author{
\IEEEauthorblockN{Jiahao Zhang}
\IEEEauthorblockA{Vanderbilt University \\ jiahao.zhang@vanderbilt.edu}
\and
\IEEEauthorblockN{Yifan Zhang}
\IEEEauthorblockA{Vanderbilt University \\ yifan.zhang.2@vanderbilt.edu}
\and
\IEEEauthorblockN{Yu Huang}
\IEEEauthorblockA{Vanderbilt University \\ yu.huang@vanderbilt.edu}
}

\maketitle

\begin{abstract}

Resolving a real software issue with a large language model (LLM) agent is a long repair episode, often tens to hundreds of steps spanning exploration, hypothesis, implementation, and verification.
Success depends on both the base model's local reasoning and the agent's ability to maintain an evolving plan and remember observations across phases.
Existing repository-level agents typically strengthen planning or memory in isolation, leaving long trajectories vulnerable to stale evidence, repeated failed edits, and verification inferred from the agent's own claims instead of execution evidence.
We present PMCoder, an issue-resolution agent that couples a hierarchical phase planner with episodic memory.
The coupling is bidirectional: the current plan phase conditions memory retrieval, while memory-derived trajectory statistics inform stuck detection and replanning.
When available, issue-reproduction verdicts ground verification progress in execution evidence rather than self-reported completion.
On SWE-bench Verified, PMCoder resolves an average of $25$ more cases ($+5.0$pp) than a harness-matched baseline, with gains persisting even where the reproduction gate never fires.
Further Verified-500 evaluations show the same positive direction across Claude Haiku 4.5, DeepSeek-V4-Flash, and an OpenHands port, with at least $14$ additional resolved cases ($+2.8$pp).
Separately, evaluation on TerminalWorld's official sample suggests that the plan--memory substrate transfers beyond issue reports.
Ablation and trajectory analyses show where the gains come from: coupling planning and memory outperforms either component alone and reduces repeated failed actions, empty-patch exits, and context-window exhaustion.

\end{abstract}


\section{Introduction}\label{sec:intro}

An LLM agent that resolves a real software issue generally consists of four
functional modules: perception, planning, memory, and
action~\cite{sumers-coala,wang-se-agent-survey}. Over a long repair episode,
planning and memory carry the agent's evolving state, its current phase, and
what it has already tried; that state, more than the base model alone,
drives the outcome of the episode~\cite{yang2024sweagent, reflexion}. Most progress on
these agents has come from advances in individual modules.

Existing repository-level software agents primarily strengthen either planning
or memory. Planning-oriented approaches improve repository exploration,
localization, and edit planning through search and structured
reasoning~\cite{auto-code-rover,codeplan,swe-search}, while memory-oriented
approaches store and retrieve prior repair trajectories, demonstrations, and
accumulated experience~\cite{chen-swe-exp-2025,mu-experepair-2025,wong-confucius-2025}.

Designing the two in isolation is at odds with the nature of repository-level
issue resolution. Repair is a long-horizon process that alternates between
exploration, hypothesis, implementation, and verification. Observations discovered early in an episode
may become relevant only much later, while repeated failures should trigger a
revision of the current strategy. Effective repair therefore requires planning
and memory to interact: plans should determine which past information remains
relevant, and accumulated experience should determine when plans need to change.
This view echoes cognitive accounts of externalized reasoning, where plans and
notes form coupled external working state: the current plan determines what
evidence is relevant, and accumulated observations revise the plan
itself~\cite{clark-chalmers-extended-mind,risko-gilbert-offloading-2016,
hutchins-distributed-cognition-1995}.

Bidirectional coordination between planning and memory remains largely
unexplored for software repair. General-agent work suggests that planning can
benefit from retrieval over contextual memory~\cite{kagaya-rap-2024}, but this
idea does not transfer directly to repository-level repair. Repair state is
distributed across files, mutated by the agent's own edits, and tied to
verification outcomes that must be checked by execution. Recent issue-resolution memory work aligns retrieval with plan
structure but leaves plan updates independent of memory-derived trajectory
evidence~\cite{shen-subtask-memory}.
To the best of our knowledge, no repository-level repair agent both guides
memory retrieval by plan state and drives replanning from memory-derived signals.

We present \textit{PMCoder}, an issue-resolving agent that couples planning and
episodic memory as distinct modules for maintaining agent state. Its
hierarchical phase planner tracks two levels of state, the current repair phase
and active sub-task, while its episodic memory stores entries
for observations, attempts, outcomes, and file metadata and retrieves a budgeted
working set by
beam search over maximal-marginal-relevance
(MMR)~\cite{carbonell-goldstein-mmr} scores. The two
structures interact in both
directions: the planner's phase and sub-task guide memory retrieval, while
memory-derived statistics drive stuck detection and replanning. Plan and memory
context are injected through tool outputs, the channel that tool-calling models
already consume.

\begin{figure*}[!t]
\centering
\includegraphics[width=\textwidth]{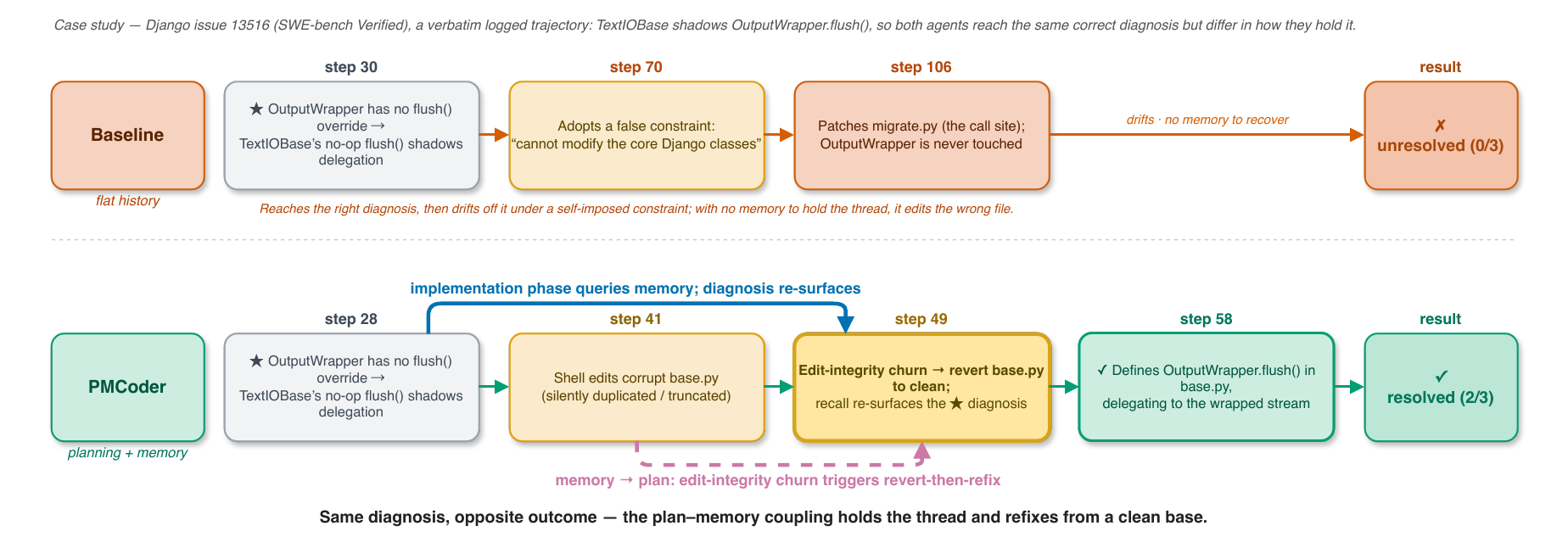}
\caption{A \texttt{django-13516} trajectory illustrating plan--memory coupling:
both agents find the same diagnosis, but PMCoder preserves and reuses it after
edit recovery while the baseline drifts to the wrong file.}
\label{fig:case-django13516}
\end{figure*}

Figure~\ref{fig:case-django13516} shows this interaction on a real
SWE-bench Verified trajectory. Both agents first reach the same diagnosis on
\texttt{django-13516}, but the baseline later drifts to the wrong file under a
self-imposed constraint. PMCoder instead re-surfaces the earlier diagnosis when
the plan returns to implementation, and memory-derived edit signals trigger
recovery from a corrupted edit state. The example illustrates the role of
plan--memory coupling: relevant evidence remains retrievable at the right phase,
and repeated failed work can change the plan rather than accumulate; we return
to this trajectory in Section~\ref{sec:results-rq2}.
This coupling relies on a shared notion of progress: the planner must know when
to advance, and memory-derived recovery signals must be attached to a trajectory
state that the agent can trust. PMCoder therefore grounds
verification-phase transitions in issue-reproduction outcomes rather than in
self-reported completion signals, which are often
unreliable~\cite{huang-self-correct-2024,gou-critic-2024}. These checks serve
the coupling by keeping the shared plan--memory state accountable to execution
evidence as the two structures exchange context and recovery signals.

We evaluate PMCoder against a harness-matched baseline agent~\cite{yang2024sweagent} on SWE-bench
Verified~\cite{swe-bench,swe-bench-verified}. The baseline has no planner, episodic memory, or execution-grounding gates, so
the primary comparison tests the complete PMCoder design. Because served FP8 models at
temperature $0$ are not fully run-reproducible, we repeat the evaluation three
times and assess significance at the instance level. Across
runs, PMCoder resolves $167.3/500$ issues ($33.5\%$) versus $142.3/500$
($28.5\%$) for the baseline on Qwen3-Coder-30B, an average improvement of
$+25.0$ resolved issues ($+5.0$pp), with a per-instance cluster-bootstrap
$95\%$ confidence interval of $[+14.3, +35.7]$ ($p < 0.001$). A repeated
$2\times2$ component ablation further isolates planning, memory, and their
interaction: the plan--memory interaction contributes $+10.3$ resolved
instances ($+2.1$pp; $F(1,8)=10.92$, $p=0.011$;
Section~\ref{sec:ablation}). A behavioral mechanism analysis shows fewer failed
action recurrences, empty-patch exits, and context-window failures, including
on instances where no reproduction script exists. Additional evaluations across other
models, a framework port, and a terminal-task benchmark provide generality
evidence beyond the headline configuration.

This paper makes the following contributions:
\begin{itemize}

\item \textbf{Design insight.}
We identify bidirectional plan--memory coupling as a design principle for
long-horizon repository repair, where plan state guides recall and accumulated
trajectory evidence drives replanning.

\item \textbf{Approach.}
We propose \textit{PMCoder}, which couples a hierarchical phase planner and
episodic memory through bidirectional plan--memory coupling.

\item \textbf{Evaluation.}
Experiments on SWE-bench Verified show that PMCoder improves issue
resolution over a harness-matched baseline. Repeated ablations identify a positive plan--memory
interaction, and additional models, a framework port, and a terminal-task benchmark
provide generality evidence.

\end{itemize}

The remainder of this paper is organized as follows.
Section~\ref{sec:background} motivates reliable plan--memory state in agentic
issue resolution. Section~\ref{sec:methodology} describes the
PMCoder architecture and its co-design couplings. Section~\ref{sec:setup}
details the experimental setup. Section~\ref{sec:results} reports results for
four research questions covering effectiveness, mechanism, generality,
and a component ablation. Sections~\ref{sec:discussion}
and~\ref{sec:threats} discuss implications and threats to validity,
Section~\ref{sec:related} surveys related work, and Section~\ref{sec:conclusion}
concludes.

An anonymized implementation and replication package is available at
\url{https://doi.org/10.5281/zenodo.21035632}.

\section{Background}\label{sec:background}

\subsection{Agentic Issue Resolution on SWE-bench Verified}

An LLM issue-resolving agent runs an observe--act loop over a
repository~\cite{yang2024sweagent}. Given an issue report and a copy of the repository
at the faulty commit, it alternates between emitting shell commands and reading
their outputs until it submits a patch or exhausts a budget. This loop underlies
systems from minimal single-tool agents to frameworks such as
mini-SWE-agent~\cite{yang2024sweagent} and OpenHands~\cite{openhands}.

We evaluate on SWE-bench Verified, a human-validated set of 500 real GitHub
issues~\cite{swe-bench, swe-bench-verified}. An instance is resolved only when
the submitted patch passes the project's withheld tests.
This output-level grading matters: a noisy trajectory that ends in a correct
patch ties a clean one, while a confident trajectory that ends wrong scores
zero. The official grade ignores how the agent reached its patch, but the path
is determined by trajectory-level state the agent keeps for itself.

\subsection{Coupled Planning and Memory}

Resolving one issue often takes tens to hundreds of steps across exploration,
hypothesis formation, implementation, and verification. A flat transcript of
that episode neither fits the context window nor stays navigable, so agents
need structured state: a plan of the current phase and sub-task~\cite{prasad-adapt-2024},
and episodic memory of what they have read, tried, and observed. Recent repair
agents also retrieve past repair experience to steer the next
attempt~\cite{chen-swe-exp-2025, mu-experepair-2025, wong-confucius-2025}.

Prior SWE repair agents commonly treat planning and memory as separate
capabilities. Planners operate over the current history; memory systems retrieve
against the issue text or accumulated experience, rather than being explicitly
conditioned on the current phase. The long-horizon setting makes these states
mutually dependent: the phase decides what is worth recalling, and accumulated
evidence decides whether to continue, recover, or replan. Yet issue-resolution
agents largely leave this coupling implicit.

The resulting structured state governs the episode: it affects what the model
sees next, when the agent stops, and which patch is submitted. Its usefulness
therefore depends on the signals used to advance it.

\subsection{The Success-Signal Problem}

Inside an agentic repair trajectory, progress signals are sparse. With the
official oracle withheld, the agent mainly sees execution feedback from its
commands and textual claims about its own progress. Neither reliably implies
resolution. Return codes are coarse: for example, \texttt{ls}, \texttt{grep},
and a diagnostic print statement can all exit~0, making success difficult to
infer from return status alone.
Self-written tests are also unreliable success signals, since the model that
wrote the fix can also write an assertion that matches the patched behavior
rather than the issue requirement~\cite{qi-patch-plausibility,
smith-overfitting-2015}. A free-text ``the fix is verified'' is a claim, not an
observation.

Self-reported verification is a textual success claim without an
execution-backed check of the issue requirement. In a small motivating audit, we
examined the first nine Qwen pilot trajectories we encountered whose final turn
made such a claim, and observed two fabricated-verification patterns: narrative
prints that echoed success without running a relevant test, and self-written
always-pass tests that checked the patched code's own behavior.
This is consistent with broader reports of specification gaming and reward
hacking~\cite{pan-reward-misspec-2022, skalse-reward-hacking-2022,
denison-reward-tampering-2024}.
The audit is not a prevalence estimate, but it shows why self-report is unsafe
as a control signal: an agent that advances its plan state on self-reported
success can turn claims into memory and recovery decisions. Large-scale agent
evaluations point in the same direction, with one study observing a $22\%$ task
success rate alongside a $77\%$ predicted success
rate~\cite{kaddour-overconfidence-2026}.

Execution-based validation in automated program repair traditionally judges
finished patches~\cite{genprog}. Agentic resolution exposes a different surface:
the in-episode plan state that steers retrieval, recovery, and stopping.
Whether ``exploration is done'' or ``this sub-task is verified'' is not graded
by SWE-bench, yet it determines the path to the final patch.
PMCoder targets this process-level gap by making planning and memory share an
explicit trajectory state, whose verification updates are grounded in behavioral
issue-reproduction verdicts rather than self-report; Section~\ref{sec:related}
contrasts this supporting use of execution evidence with output-level
program-repair oracles.

\section{Methodology}\label{sec:methodology}

\label{sec:method}

\subsection{Overview}
\label{sec:method:overview}

\begin{figure*}[t]
  \centering
  \includegraphics[width=\textwidth]{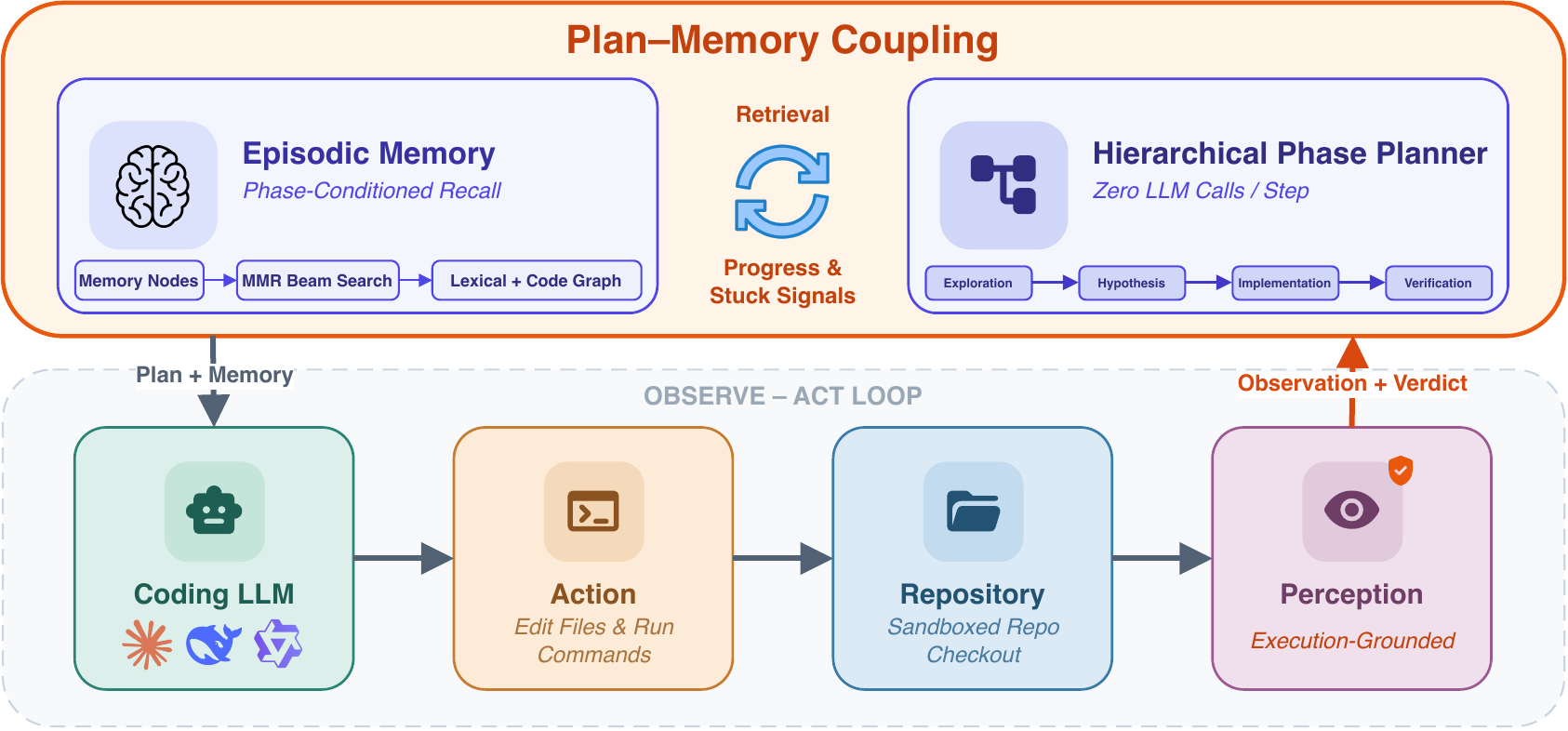}
  \caption{PMCoder couples episodic memory with hierarchical planning
  inside the observe--act loop.}
  \label{fig:architecture}
\end{figure*}

PMCoder keeps the observe--act loop of Section~\ref{sec:background} intact: the
agent still operates on a sandboxed copy of the target repository and submits a
final patch. On top of this loop it adds two coupled state structures, supported
by delivery through tool results and execution-grounded verification updates
(Figure~\ref{fig:architecture}). The two structures are a hierarchical phase
planner that maintains explicit plan state and an episodic memory that retrieves
a budgeted working set. Their bidirectional coupling is the center of the
design: plan state guides retrieval, while memory-derived evidence drives stuck
detection and replanning. Delivering state through tool results preserves the
message history, and execution grounding keeps verification-phase updates tied
to issue-reproduction behavior. The four boundary-crossing couplings are
summarized at the end of the section.

Two invariants hold throughout. First, PMCoder delivers added state through
channels a tool-calling model already expects, without reordering or rewriting
prior turns. Second, the added signals are advisory for the model and internal
to PMCoder: they may update internal plan state, mark a verification sub-task
incomplete, or suggest recovery, but they do not block actions, directly revert
edits, or participate in official SWE-bench grading. The benchmark grader sees
only the submitted patch, while RQ2 measures whether these advisory signals
change observable repair behavior.

\subsection{Hierarchical Phase Planner}
\label{sec:method:planner}

\begin{figure}[t]
  \centering
  \includegraphics[width=\columnwidth]{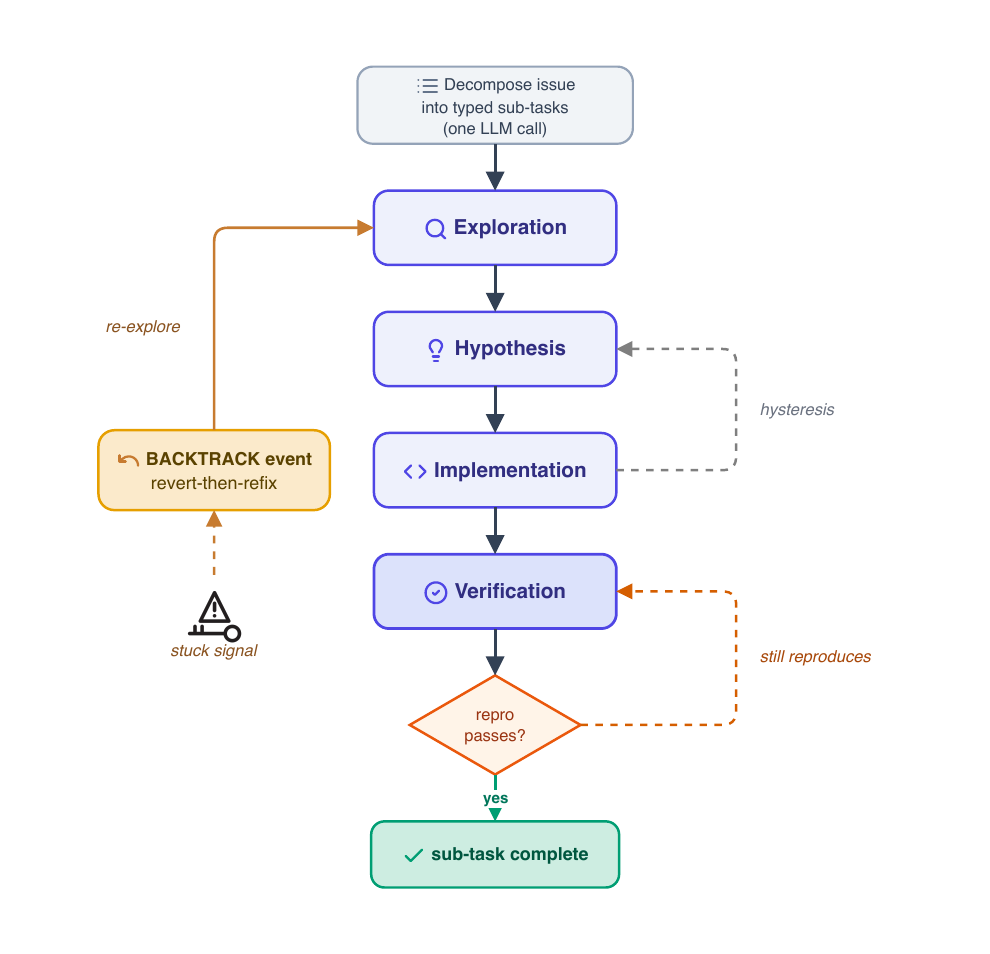}
  \caption{The phase planner advances sub-tasks through four phases with
  hysteresis, backtracking, and execution-grounded verification.}
  \label{fig:planner-sm}
\end{figure}

\paragraph{Plan state.}
The planner maintains plan state as a deterministic state machine
(Figure~\ref{fig:planner-sm}) over four
progress phases (\textsc{exploration}, \textsc{hypothesis},
\textsc{implementation}, \textsc{verification}) ordered by a scalar rank, plus a
\textsc{backtrack} event held outside the ordering, since backtracking is
a recovery transition. A single
start-of-task LLM call decomposes the issue into typed sub-tasks,
each tagged with its target phase, such as a verification sub-task to re-run
the failing test; a parse failure falls back to a fixed skeleton instead of
failing silently. This is the planner's only LLM call. Per-step phase detection,
sub-task bookkeeping, backtracking, and replanning are deterministic.

\paragraph{Phase detection.}
Each step, a zero-cost rule-based detector proposes a phase from the last shell
command and the model's stated reasoning. The detector normalizes shell
wrappers and then applies ordered rules over four signal classes:
file-inspection behavior and exploratory language map to \textsc{exploration};
diagnostic commands or root-cause language map to \textsc{hypothesis};
file-mutating actions map to \textsc{implementation}; and test execution or
explicit verification language maps to \textsc{verification}. Ordered rules
prioritize mutating actions over inspection and treat diagnostic one-off
executions as hypothesis-building. Because the detector is cheap but noisy, the
planner accepts forward phase moves immediately but requires repeated evidence
before backward or lateral moves take effect.

\paragraph{Backtracking and replanning.}
After a minimum-step floor, the planner declares the agent stuck on any of
several signals: persistently nonzero return codes, a single file edited too many
times, file reads saturated without edits, or one normalized action repeated past
a threshold. The memory subsystem computes several of these signals and passes
them to the planner as aggregate statistics. When a backtrack is triggered, the
planner marks the active sub-task as failed. It then pushes recovery sub-tasks
onto a goal stack, including execution grounding's revert-then-refix template,
which restores the edited files and re-fixes from a clean base. A cooldown and a depth bound prevent long, repetitive trajectories
from inducing bursts of repeated replanning.

\paragraph{Completion and the exported signal.}
A non-terminal sub-task completes once the detected phase has advanced past its
own phase with minimal supporting evidence. A terminal \textsc{verification}
sub-task has no higher phase to enter, so it completes only on a successful
verification step, using the completion predicate hardened by execution
grounding. Each step, the planner exports a compact signal: the
current phase, the active sub-task, progress counts, the most-touched files, a
per-phase token budget, and the backtrack flag. This exported signal is what
conditions memory retrieval.

\subsection{Episodic Memory}
\label{sec:method:memory}

\paragraph{Memory nodes.}
Each memory node stores not only text but also structured
metadata about the episode event: its message role, recency identifier,
compressed content view, summary, whether the preceding command edited files,
and which files it touched. Every message is mirrored into such a node, with
the identifier equal to its message position, so recency and truncation follow
the conversation order. For an observation node, the
edit and file-touch metadata are read from the executed command in the preceding
assistant message, so memory is grounded in what the agent ran, not in what the
model reports running. The graph stores raw observations only; injected context
mutates the live conversation, never the graph, so retrieval cannot re-ingest
its own output.

\paragraph{Budgeted MMR retrieval.}
Retrieval selects a working set under a token budget using the
maximal-marginal-relevance (MMR) selection criterion~\cite{carbonell-goldstein-mmr}.
A fixed core is always kept: the task statement, the recent tail, and the anchor
node, which holds the current query focus seeded from the active sub-task
keywords. A beam search adds the rest, scoring each candidate $v$ by marginal gain
\[
  g(v) \;=\; \lambda \cdot \mathit{rel}(v) \;-\; (1-\lambda) \cdot
  \max_{u \in S} \mathit{sim}(v, u),
\]
where $\lambda$ is the relevance-diversity tradeoff, $S$ is the selected set,
and $\mathit{sim}$ is IDF-weighted Jaccard overlap.
Relevance fuses two signals,
\[
  \mathit{rel}(v) = w_c \cdot \mathit{lex}(v) + w_g \cdot \mathit{graph}(v).
\]
Here, $w_c$ and $w_g$ are the lexical and graph weights. $\mathit{lex}$ is an
IDF-weighted lexical score over anchor words, while $\mathit{graph}$ scores
proximity in a code-structure file graph built
incrementally from the trajectory (same-file co-occurrence, Python import edges,
and trajectory adjacency). Import edges are extracted from observed Python
snippets by AST parsing, with a conservative regex fallback when parsing fails.
Fusing code structure on top of lexical match makes retrieval structure-aware
rather than purely textual. Section~\ref{sec:setup-hyperparams}
reports the fixed retrieval parameters used in the final evaluation.

\paragraph{Phase-conditioned retrieval policy.}
A memory controller maps the planner's phase to retrieval parameters, so the
working set adapts to the agent's current phase. Exploration and backtracking use a
large, diversity-favoring budget, retrieving broadly while the agent orients or recovers;
implementation draws a small, graph-weighted budget focused tightly on the edit
site. This phase-to-policy map is the plan$\rightarrow$memory direction of the
coupling; the concrete budgets and weights are given in
Section~\ref{sec:setup-hyperparams}.

\subsection{In-Distribution Injection Channel}
\label{sec:method:channel}

\begin{figure}[t]
  \centering
  \includegraphics[width=\columnwidth]{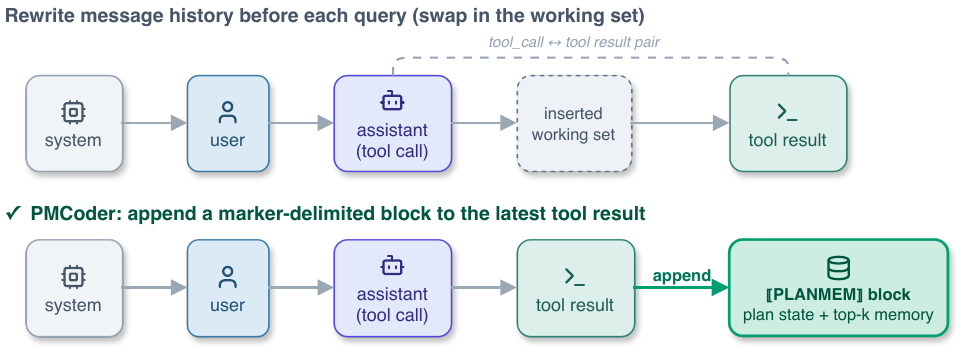}
  \caption{PMCoder delivers plan--memory state by appending a marker-delimited
  block to the latest tool result.}
  \label{fig:injection}
\end{figure}

Retrieved memory and plan state must still reach the model. A natural approach,
used by working-memory management agents, rewrites the message history before
each query, for example by replacing past observations with summaries or
swapping the raw trajectory for a selected working set~\cite{hu-hiagent-2025}.
In tool-call mode this needs care: the API pairs each assistant tool call with
the tool-result message that answers it. Reordering those messages separates the
pair and can push the request off-distribution.
PMCoder therefore leaves history
untouched and instead appends a compact, marker-delimited block to the
most-recent tool-result message (Figure~\ref{fig:injection}): a plan-state
header (phase, active sub-task,
progress) and the top retrieved nodes, bounded by the injection budget.
To the model, this arrives as part of the environment's reply, the channel a
tool-calling model is trained to read, so each request stays in-distribution; our
pilot design observation was that
tool-result placement was more stable than a separate user-role reminder. Two
filters bound growth: injection fires only when the plan state changes, and a
novelty filter drops already-surfaced nodes in favor of a short prescriptive
line, such as a one-line summary of the failed sub-task or a file-churn count.

\subsection{Execution Grounding}
\label{sec:method:grounding}

The planner and memory create a shared plan--memory state, and the injection
channel delivers that state without rewriting history. The remaining question is
how this state advances through verification. Without grounding, verification
depends on coarse return codes and self-reported success, the weak signals
discussed in Section~\ref{sec:background}. The grounding layer replaces them
with a behavioral check that keeps the coupled state tied to execution evidence
without changing what the agent may do.

For each instance with a validated issue-reproduction script
(Section~\ref{sec:setup-repro}), the grounding layer reruns the script after a
file-mutating edit. The verdict is pass if the issue no longer reproduces and
fail if it still does; instances without a validated script use the ungated
completion predicate while retaining the same plan--memory substrate (the coupled
planner and memory). The verdict
has two consumers. As advisory text on the current tool result, it tells the
model whether the once-failing reproduction now passes or still fails. As an
update to internal plan state, it hardens the completion predicate. In a
verification step, a zero return code counts as evidence only if the latest
verdict is not a fail. A terminal verification sub-task therefore cannot complete
while the issue still reproduces.

Edit integrity and recovery provide a lighter grounding signal for corrupted
edits. After a mutating edit to a Python file, a compile check appends a
corrective note when the edit leaves the file unparsable, identifying the
corrupted state before the agent compounds the error with further edits. When a
sub-task fails amid heavily edited files, the planner's recovery template first
restores them to their committed state, then re-examines the failure and refixes
from the clean base. This follows the delta-debugging principle of restarting
from a known-good state instead of patching a corrupted
one~\cite{zeller-hildebrandt-delta-debugging}.

Every grounded signal remains advisory, and the harness, which alone assigns the
resolution label, never reads the verdict. PMCoder's
distinction lies in where the verdict lands: on the agent's evolving plan state
mid-trajectory, the same state that conditions memory retrieval and completion
checks, rather than on a finished patch.

\subsection{Co-Design Couplings}
\label{sec:method:couplings}

The preceding subsections define the planner, memory, delivery channel, and
grounding layer. The design depends on four explicit couplings between them,
each crossing a component boundary and together realizing the bidirectional
plan--memory design.

\paragraph{Phase $\rightarrow$ retrieval.}
The planner's phase selects the retrieval budget, diversity pressure, and graph
weights, reducing irrelevant or stale recall.
\paragraph{Sub-task $\rightarrow$ retrieval.}
Keywords from the active sub-task enter the retrieval anchor, countering
goal-agnostic recency bias.
\paragraph{Memory statistics $\rightarrow$ plan.}
Edit counts, read saturation, and repeated actions inform stuck detection and
replanning, reducing repeated failed work.
\paragraph{Verdict $\rightarrow$ plan.}
Issue-reproduction pass/fail verdicts update verification-phase state, reducing
premature verification.

The first three couplings run between planner and memory: plan state conditions
what memory surfaces, while memory statistics reshape the plan. The fourth ties
execution evidence into that same plan state. These dependencies make the
planner and memory a co-designed control structure rather than two independent
add-ons.

\section{Experimental Setup}\label{sec:setup}

\subsection{Benchmark and Metric}
\label{sec:setup-benchmark}

We evaluate on SWE-bench Verified, the human-validated 500-instance slice of
SWE-bench~\cite{swe-bench, swe-bench-verified}, abbreviated
\emph{Verified-500}. Each instance is graded only by the official SWE-bench
harness: a submitted patch is \emph{resolved} if it passes the instance's
fail-to-pass and pass-to-pass tests. All arms use the full 500-instance
denominator; errored runs and non-applicable patches count as unresolved.

\subsection{Research Questions}
\label{sec:setup-rqs}

Our evaluation answers four questions:
\begin{itemize}
\raggedright
\item \textbf{RQ1 (Effectiveness):} Does PMCoder resolve software issues more
effectively?
\item \textbf{RQ2 (Mechanism):} What trajectory evidence explains PMCoder's
gains?
\item \textbf{RQ3 (Generality):} Does the effect persist beyond the headline
configuration?
\item \textbf{RQ4 (Component ablation):} Does coupled planning and memory
outperform either component alone?
\end{itemize}

\subsection{Agents and Arms}
\label{sec:setup-arms}

All agents run in the same SWE-bench Docker environments, execute shell commands
over the target repository, and submit only a final patch for official grading. The
baseline is the mini-SWE-agent loop with PMCoder's planner, memory,
injection, and grounding components disabled. The headline baseline and
PMCoder arms use three runs. RQ4 uses the same protocol for plan-only and
memory-only variants. RQ3 includes DeepSeek, Claude Haiku,
OpenHands, and TerminalWorld's official 20-task sample.

\subsection{Implementation and Model Serving}
\label{sec:setup-implementation}

The headline Qwen3-Coder model~\cite{qwen3technicalreport} is served through
vLLM~\cite{kwon-vllm} and accessed through LiteLLM tool calls.
Table~\ref{tab:setup-serving} gives the shared runtime configuration and RQ3
model identifiers.

\begin{table}[t]
\caption{Serving and runtime configuration.}
\label{tab:setup-serving}
\centering
\begin{tabular}{p{0.30\columnwidth}p{0.58\columnwidth}}
\toprule
Setting & Value \\
\midrule
Serving & vLLM, FP8, LiteLLM tool calls \\
Headline model & \texttt{Qwen3-Coder-30B-A3B-Instruct-FP8} \\
DeepSeek model & \texttt{deepseek/deepseek-v4-flash} \\
Claude model & \texttt{anthropic/claude-haiku-4-5} \\
Sampling & temperature 0.0 \\
Context cap & 65,536 tokens (vLLM) \\
Generation cap & no client-side cap \\
Step budget & 250 assistant turns \\
Cost cap & \$3.0 per instance \\
Shell timeout & 60 seconds \\
Retry policy & 10 attempts, exponential backoff \\
\bottomrule
\end{tabular}
\end{table}

\subsection{PMCoder Hyperparameters}
\label{sec:setup-hyperparams}

Table~\ref{tab:setup-hyperparams} lists the behavioral parameters, fixed from
pilot runs before the final Verified-500 evaluations. These values are the
default PMCoder configuration used for all reported runs; they keep retrieval
within the context budget, debounce noisy phase changes, and bound recovery
frequency. Phase budgets apply when the planner is active; memory-only uses the
16k default.

\begin{table}[t]
\caption{Fixed PMCoder hyperparameters.}
\label{tab:setup-hyperparams}
\centering
\begin{tabular}{p{0.33\columnwidth}p{0.25\columnwidth}p{0.30\columnwidth}}
\toprule
Parameter & Value & Role \\
\midrule
MMR $\lambda$ & 0.7 & relevance/diversity \\
Lexical/graph weights & 0.5/0.5 & base relevance mix \\
Memory budget & 16k tokens & default retrieval budget \\
Node cap & 4k chars & per-memory node cap \\
Phase budgets & 20k/16k/12k/16k/20k & expl/hyp/impl/verif/back \\
Phase hysteresis & 2 steps & debounce \\
Min backtrack step & 10 & suppress early recovery \\
Backtrack cooldown & 8 steps & rate-limit replanning \\
Goal-stack depth & 16 & bound recovery stack \\
Failed returns & 5 & stuck threshold \\
Repeated edits & 6 & stuck threshold \\
Read saturation & 15 & stuck threshold \\
Repeated action & 8 & stuck threshold \\
Repro checks & 4 & per-episode cap \\
Repro timeout & 180s & per-check timeout \\
\bottomrule
\end{tabular}
\end{table}

\subsection{Issue-Reproduction Scripts}
\label{sec:setup-repro}

PMCoder's execution grounding uses per-instance issue-reproduction scripts
extracted offline from issue text alone by the same Qwen3-Coder model used in
the headline system. The prompt asks for a self-contained bash or Python
reproducer and does not expose the gold patch or hidden tests.

Candidates are validated on the unmodified repository in the same Docker
environment used by the agent and admitted only if they reproduce the pre-edit
failure within the configured timeout. Otherwise the verdict is undefined and
PMCoder falls back to the ungated completion predicate.

During a run, reproduction checks use their own timeout
(Table~\ref{tab:setup-hyperparams}) and verdicts are cached by (instance, patch
signature). These verdicts only affect internal verification-phase state and
recovery choices; all reported resolution numbers come from the official
SWE-bench harness.

\subsection{Statistical Protocol}
\label{sec:setup-statistics}

The headline Qwen comparison uses three runs per arm because FP8 serving is not
bit-reproducible even at temperature~0. We therefore treat each run as a
replicate and aggregate the headline comparison at the instance level.

For the headline confidence interval, we use an instance-level cluster
bootstrap. Each replicate samples 500 instance identifiers with replacement,
averages each sampled instance's resolved indicator across the three runs per
arm, and computes the mean paired difference. The reported 95\% confidence
interval is the percentile interval; the $p$ value is the two-sided bootstrap
tail probability around zero.

As a complementary paired test, McNemar tests compare matched runs and count
instances solved by only one arm. RQ4 fits a standard run-level $2\times2$
factorial model with planning, memory, and their
interaction as fixed effects.
RQ2 trajectory signatures are averaged per instance before paired intervals are
formed. The difficulty breakdown is reported as a supporting analysis; RQ3
evaluations are costlier single-run supporting generality probes. All grading uses the
official Docker harness; errored agent
trajectories are retained and scored unresolved.

\section{Results}\label{sec:results}

We organize the results around the four research questions stated in
Section~\ref{sec:setup}: effectiveness (RQ1), trajectory
evidence explaining the gains (RQ2), generality beyond the headline
configuration (RQ3), and a component ablation of coupled planning and memory
(RQ4). Unless noted
otherwise, all numbers are resolved-instance counts and resolve rates on
Verified-500 under the official SWE-bench harness~\cite{swe-bench,
swe-bench-verified}.

\subsection{RQ1: Does PMCoder resolve software issues more effectively?}

\begin{table}[t]
\caption{Resolved counts and rates on Verified-500 across three runs.}
\label{tab:rq1-main}
\centering
\begin{tabular}{lccccc}
\toprule
Arm & Run 1 & Run 2 & Run 3 & Mean & Mean rate \\
\midrule
Baseline       & 139 & 144 & 144 & 142.3 & 28.5\% \\
PMCoder   & 170 & 164 & 168 & 167.3 & 33.5\% \\
\midrule
Improvement    & $+31$ & $+20$ & $+24$ & $+25.0$ & $+5.0$pp \\
\bottomrule
\end{tabular}
\end{table}

Table~\ref{tab:rq1-main} reports three runs per arm under the final
configuration. PMCoder resolves a mean of $167.3$ instances ($33.5\%$,
range $164$--$170$) against the baseline agent's $142.3$ ($28.5\%$, range
$139$--$144$), a mean gap of $+25.0$ instances or $+5.0$ percentage points.
The two arms separate completely: the weakest PMCoder run, $164$
resolved instances ($32.8\%$), exceeds the strongest baseline run, $144$
($28.8\%$), by $20$ instances.

Using the repeated-run protocol in Section~\ref{sec:setup-statistics}, the
cluster bootstrap estimates the gain at $+25.0$ with a $95\%$ confidence
interval of $[+14.3, +35.7]$ ($p < 0.001$), and the three trial-paired McNemar tests
are each significant ($p$ between $0.002$ and $0.033$). The improvement is
therefore statistically significant and robust to serving noise.

To understand where this aggregate gain lands, we stratify the first-seed paired run by
the SWE-bench Verified difficulty estimate (OpenAI's human fix-time annotation)
\cite{swe-bench-verified} (Table~\ref{tab:rq1-difficulty}). On the two tiers that
the base model can solve, PMCoder raises the resolve rate by a
near-constant margin: $+6.2$ percentage points on the $<$15\,min tier
($89 \to 101$) and $+6.5$ points on the 15\,min--1\,h tier ($50 \to 67$).
The long-fix-time tail behaves differently: across the $45$ instances
estimated to take more than an hour,
the baseline resolves none and PMCoder resolves only two. Because the same
base model under the baseline agent already resolves zero of these, we read
the tail as a base-model capability ceiling. PMCoder therefore mainly reduces
state-management failures within the base model's capability frontier.

\begin{table}[t]
\caption{Resolved counts by SWE-bench Verified difficulty.}
\label{tab:rq1-difficulty}
\centering
\begin{tabular}{lcccc}
\toprule
Difficulty & $n$ & Baseline & PMCoder & Improvement \\
\midrule
$<$15\,min    & 194 & 89 & 101 & $+12$ \\
15\,min--1\,h & 261 & 50 & 67 & $+17$ \\
1--4\,h       & 42  & 0  & 2  & $+2$  \\
$>$4\,h       & 3   & 0  & 0  & $0$   \\
\bottomrule
\end{tabular}
\end{table}

\subsection{RQ2: What trajectory evidence explains PMCoder's gains?}
\label{sec:results-rq2}

\paragraph{Behavioral signatures of state management}
If the plan--memory coupling helps by keeping a long repair episode aligned with
accumulated evidence, the effect should be visible in \emph{how} the agent
behaves, not only in resolved counts. Table~\ref{tab:rq2-behavior} reports four
trace-derived signatures over the three RQ1 runs per arm. They capture repeated
normalized commands after failure, terminal give-up or empty-patch outcomes,
terminal context-window exhaustion, and revert-then-refix recoveries. The first
three signatures are failure rates, where lower is better; the last is a
recovery count, where higher indicates more successful recovery attempts. Ratios
are PMCoder divided by baseline.

These trajectory-level measures are computed per trajectory and then averaged
per instance; the rate-based measures normalize for PMCoder's slightly longer
episodes ($51.9$ vs.\ $47.3$ steps). PMCoder re-issues failed commands about
half as often, gives up with empty or no-op patches about one third as often,
and exhausts its context window less often. It also shows more
revert-then-refix recoveries, indicating that the agent more often returns to a
clean state after a failed edit.

\begin{table}[t]
\caption{Trajectory-level state-management signatures over the three RQ1 runs per arm.}
\label{tab:rq2-behavior}
\centering
\small
\setlength{\tabcolsep}{3pt}
\renewcommand{\arraystretch}{1.06}
\begin{tabular}{@{}p{0.48\columnwidth}ccc@{}}
\toprule
Signature & Baseline & PMCoder & ratio \\
\midrule
\multicolumn{4}{@{}l}{\emph{Failure rates (lower is better)}} \\
Failed-action recurrence rate    & 0.0137 & 0.0069 & 0.50$\times$ \\
Give-up / empty-patch rate       & 8.3\%  & 2.7\%  & 0.33$\times$ \\
Context-window exhaustion rate   & 6.7\%  & 3.0\%  & 0.45$\times$ \\
\midrule
\multicolumn{4}{@{}l}{\emph{Recovery count (higher is better)}} \\
Revert-then-refix recoveries     & 2.89   & 4.23   & 1.46$\times$ \\
\bottomrule
\end{tabular}
\end{table}

The final row reports a positive recovery behavior rather than a failure rate:
revert-then-refix recoveries count trajectories where PMCoder returns to a clean
edit state before continuing. The Django case at the end of this subsection
illustrates this pattern in detail.

\paragraph{Predictive validity of the repro verdict}
In the Docker-validated verdict subset, the post-edit verdict is informative.
It passes for $18/20$
resolved trajectories ($90\%$) but only $5/18$ unresolved trajectories
($28\%$). The verdict is therefore a strong in-flight progress signal:
passing verdicts concentrate among trajectories that ultimately resolve, while
failing verdicts concentrate among trajectories that still need repair.

\paragraph{Armed and unarmed strata}
In the same first-seed paired run, we also separate instances by whether the
issue-reproduction gate can fire. On
unarmed instances, where no validated repro script is available and the gate is
inert, PMCoder improves from $94/315$ ($29.8\%$) to $106/315$
($33.7\%$), a gain of $+12$ cases ($+3.8$pp). On armed instances, it improves
from $45/185$ ($24.3\%$) to $64/185$ ($34.6\%$), a gain of $+19$ cases
($+10.3$pp). The larger armed gain shows the value of execution-grounded
verification, while the unarmed gain shows that the plan--memory substrate
continues to help when grounding never fires.

\paragraph{Qualitative example}
We trace one Verified instance that PMCoder resolves and the baseline agent
does not, \texttt{django-13516} (Figure~\ref{fig:case-django13516}).
Django's management-command output wrapper, \texttt{OutputWrapper}, subclasses
\texttt{io.TextIOBase} and delegates missing attributes to the wrapped stream.
However, \texttt{TextIOBase} already supplies a no-op \texttt{flush()}, so
\texttt{flush()} calls never reach the wrapped stream and \texttt{migrate}
progress remains buffered. The gold fix is to define
\texttt{OutputWrapper.flush()} in \texttt{base.py} and delegate to the wrapped
stream.

Both agents identify this root cause. The baseline later adopts a false
constraint, ``cannot modify the core Django classes'', and patches
\texttt{migrate.py}, which never touches the faulty wrapper. PMCoder also
briefly corrupts \texttt{base.py}, but the next injected tool result combines
two state-management signals: edit-integrity statistics trigger a revert to a
clean file, and memory recall re-surfaces the earlier \texttt{OutputWrapper}
diagnosis. With the diagnosis restored to context, PMCoder adds \texttt{flush()}
at the gold location and resolves the issue. This trajectory instantiates two
of the signatures in Table~\ref{tab:rq2-behavior}: PMCoder stops re-issuing the
corrupting edit and keeps its earlier diagnosis in view. PMCoder resolves the
instance in two of three runs, while the baseline fails it in all three.

\subsection{RQ3: Does the effect persist beyond the headline configuration?}

RQ3 uses single-run generality checks beyond the headline Qwen configuration.

\paragraph{Across LLMs}
\begin{table}[t]
\caption{Cross-model resolved counts and rates on Verified-500.}
\label{tab:rq3-models}
\centering
\small
\setlength{\tabcolsep}{3pt}
\begin{tabular}{@{}lccc@{}}
\toprule
Model & Baseline & PMCoder & Improvement \\
\midrule
DeepSeek-V4-Flash & 341 (68.2\%) & 357 (71.4\%) & $+16$ ($+3.2$pp) \\
Claude Haiku 4.5 & 313 (62.6\%) & 327 (65.4\%) & $+14$ ($+2.8$pp) \\
\bottomrule
\end{tabular}
\end{table}

Table~\ref{tab:rq3-models} reports cross-model checks using the served model
identifiers in Table~\ref{tab:setup-serving}. PMCoder improves over both matched
baselines: DeepSeek-V4-Flash gains $+16$ instances ($+3.2$pp), and
Claude Haiku 4.5 gains $+14$ instances ($+2.8$pp). These smaller but positive
gains are consistent with PMCoder addressing state loss rather than replacing
the base model's reasoning.

\paragraph{Framework port}
\begin{table}[t]
\caption{OpenHands framework-port results on Verified-500 with Qwen.}
\label{tab:rq3-framework}
\centering
\begin{tabular}{lccc}
\toprule
Agent & Solved & Rate & Improvement \\
\midrule
OpenHands & 146 & 29.2\% & --- \\
PMCoder & 169 & 33.8\% & $+23$ ($+4.6$pp) \\
\bottomrule
\end{tabular}
\end{table}

Table~\ref{tab:rq3-framework} adds a framework dimension by porting PMCoder's
plan--memory substrate to OpenHands V1 SDK~\cite{openhands} while keeping the
Qwen model, step budget, and official harness matched. PMCoder improves from
$146/500$ solved instances ($29.2\%$) to $169/500$ ($33.8\%$), a gain of $+23$
cases ($+4.6$pp), showing that the effect is not tied to the original
mini-SWE-agent loop.

\paragraph{Terminal-task benchmark}
On TerminalWorld~\cite{terminalworld}, using the benchmark's official 20-task
sample from its human-verified subset, PMCoder resolves $7/20$ tasks ($35\%$)
versus $5/20$ ($25\%$) for the matched Qwen baseline. Because these tasks are
not issue reports with reproduction scripts, the result exercises the
plan--memory substrate outside SWE-bench-style repair.

\subsection{RQ4: Does coupled planning and memory outperform either component alone?}
\label{sec:ablation}

RQ4 tests whether the integrated plan--memory design outperforms either
component in isolation. We run a
$2\times2$ component ablation under the headline Qwen setting, comparing
baseline, plan-only, memory-only, and the full plan+memory system under the
identical harness, model, and budget, toggling only the plan and memory
components (Table~\ref{tab:rq4-ablation}).

\begin{table}[t]
\caption{Component ablation on Verified-500 using three-run means.}
\label{tab:rq4-ablation}
\centering
\begin{tabular}{lccc}
\toprule
Configuration & Mean & Rate & Improvement \\
\midrule
Baseline      & 142.33 & 28.5\% & --- \\
Plan-only     & 148.67 & 29.7\% & $+6.3$ ($+1.3$pp) \\
Memory-only   & 150.67 & 30.1\% & $+8.3$ ($+1.7$pp) \\
Plan $+$ memory & 167.33 & 33.5\% & $+25.0$ ($+5.0$pp) \\
\bottomrule
\end{tabular}
\par\vspace{2pt}
\parbox{\columnwidth}{\raggedright\scriptsize Means are over three runs; effect
sizes and the $2\times2$ factorial test are computed from unrounded resolved
counts.}
\end{table}

\begin{figure}[t]
  \centering
  \includegraphics[width=0.8\columnwidth]{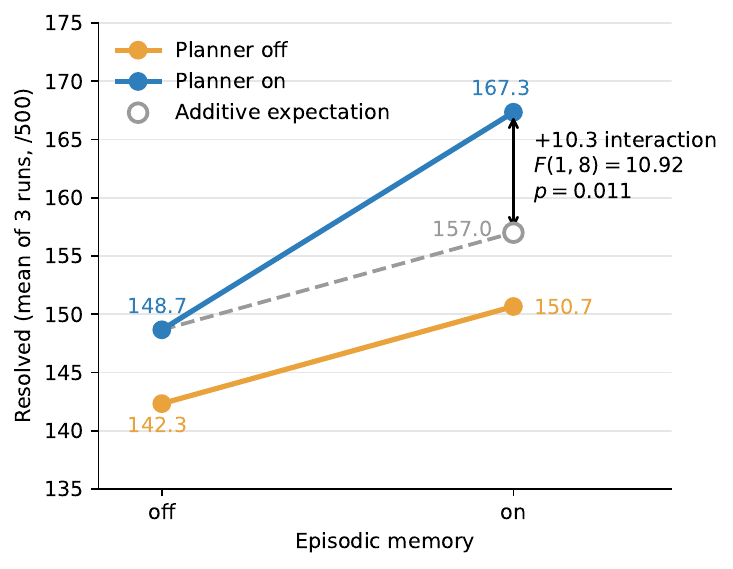}
\caption{Plan--memory interaction on Verified-500 (three-run means); the dashed
line marks the additive expectation from isolated components.}
  \label{fig:rq4-interaction}
\end{figure}

The factorial test for coupling is whether the plan+memory gain exceeds the
additive contribution of the two isolated components. We fit a balanced
$2\times2$ run-level factorial model to the resolved counts, with planning,
memory, and their interaction as fixed effects and the three repeated runs in
each cell as replicate observations. Let $\bar{Y}_{pm}$ denote the mean
resolved count with planning $p$ and memory $m$ enabled ($p,m\in\{0,1\}$). We
compute the interaction contrast as
\[
  \bar{Y}_{11}-\bar{Y}_{10}-\bar{Y}_{01}+\bar{Y}_{00}.
\]
Using the unrounded cell means, this contrast is $+10.3$ resolved instances
($+2.1$pp); Figure~\ref{fig:rq4-interaction} visualizes this as the gap between
the plan+memory point and the additive expectation. This
interaction is significant at the run
level ($F(1,8)=10.92$, $p=0.011$). Equivalently, the integrated gain over
baseline is $+25.0$ instances ($+5.0$pp), larger than the $+14.7$ instances
($+2.9$pp) expected from stacking the unrounded single-component gains. The
ablation therefore shows that coupling planning and memory performs better than
adding the two isolated components.

\section{Discussion}\label{sec:discussion}

\noindent\textbf{What PMCoder helps with.}
The gains primarily reflect better state management rather than a change in
base-model reasoning ability. The difficulty split (RQ1) concentrates the
improvement on the tiers the base model can already solve, while the hard tail
stays difficult for both arms. The trajectory signatures (RQ2) confirm the
mechanism: fewer re-issued failed actions,
empty-patch exits, and context-window failures. PMCoder therefore helps when a
fix is within reach but the episode is long enough to lose it.

\noindent\textbf{Coupling as state management.}
The central contribution is the bidirectional exchange between plan state and
memory. The ablation (RQ4) shows that the coupled system performs better than
adding the isolated planning and memory components, and the gain persists on
unarmed instances where no reproduction gate is active. Execution grounding
supports this coupling by
using execution evidence to update internal plan state, so verification progress
becomes part of the state that conditions retrieval and stuck detection.
The trajectory signatures suggest that these advisory signals affect observable
repair behavior: alongside the failure-rate drops above, PMCoder records more
revert-then-refix recoveries, the signature most directly tied to acting on an
injected recovery hint.

\noindent\textbf{Scope of the comparison.}
This mechanism study holds harness, model, budget, and tools fixed. The
primary comparison evaluates the complete PMCoder design, while RQ4 isolates
the planning and memory switches under the same setting.
Table~\ref{tab:positioning} summarizes the mechanism-level distinction. Among
prior systems, execution evidence is used for localization, patch selection, or
specification/review feedback, while recent memory systems condition either
search on retrieved experience or retrieval on sub-task structure. PMCoder
combines the two missing pieces: execution evidence updates an explicit
in-episode plan state, and that plan state is bidirectionally coupled with
episodic memory.

\begin{table*}[t]
\caption{Positioning by trajectory control, explicit memory, execution-evidence target,
and plan--memory link.}
\label{tab:positioning}
\centering
\small
\setlength{\tabcolsep}{5pt}
\renewcommand{\arraystretch}{1.08}
\begin{tabular}{@{}p{0.19\textwidth}p{0.21\textwidth}p{0.18\textwidth}p{0.21\textwidth}p{0.11\textwidth}@{}}
\toprule
System & Control & Explicit memory & Evidence updates & Link \\
\midrule
mini-SWE-agent~\cite{yang2024sweagent} & flat tool loop & -- & -- & -- \\
AutoCodeRover~\cite{auto-code-rover} & AST/code search & -- & localization / patch retry$^{\dagger}$ & -- \\
Agentless~\cite{agentless} & localize-repair-validate & -- & patch selection / validation & -- \\
SpecRover~\cite{ruan-specrover-2025} & spec inference + review loop & -- & spec / review$^{\ddagger}$ & -- \\
SWE-Exp~\cite{chen-swe-exp-2025} & SWE-Search MCTS~\cite{swe-search} & experience bank & -- & memory$\rightarrow$search \\
Shen et al.~\cite{shen-subtask-memory} & functional subtasks & subtask-aligned memory & -- & plan$\rightarrow$memory \\
\textbf{PMCoder} & \textbf{phase planner} & \textbf{episodic memory} & \textbf{plan state} & \textbf{two-way} \\
\bottomrule
\end{tabular}
\vspace{2pt}
{\raggedright\scriptsize
$^{\dagger}$AutoCodeRover uses test execution in two ways: spectrum-based
fault localization (SBFL) sharpens fault localization and context retrieval
when a test suite is available, and patch validation drives a retry loop;
neither updates an explicit in-episode plan state.
$^{\ddagger}$SpecRover runs a reproducer inside its review loop and feeds
the verdict back to the patching and reproduction agents, but the state it
updates is a natural-language specification/review, not an in-episode plan
state.\par}
\end{table*}

\noindent\textbf{Cost and deployment.}
Overhead is small relative to the effect. The planner adds one LLM call per
episode (start-of-task decomposition); all per-step planning is deterministic, and
memory retrieval is a lexical-plus-graph computation with no model call. Grounding
adds at most four cached, timeout-bounded script runs per episode, over a
reproduction table extracted once offline by the same model. In the headline
setting, this yields $+25$ resolved issues with no change to weights or serving
stack. The reproduction check is useful for mechanically reproducible issues
(crashes, traces, wrong outputs). When no such check applies, the agent falls
back to the planner--memory substrate, which the unarmed result shows still
helps. In production, the offline reproduction table could be
replaced by issue-reported steps or a failing CI job.

\section{Threats to Validity}\label{sec:threats}

\subsection{Internal Validity}

PMCoder uses Qwen3-Coder both to extract issue-reproduction scripts and to run
the headline agent. This is the lowest-cost reproducible construction path for
Verified-500; using stronger API models for script extraction would likely
improve script quality rather than advantage the local Qwen agent. We retain only
scripts that fail on the unmodified repository, use their verdicts only for
in-episode state, and grade all patches with the official SWE-bench harness.
Residual script errors may therefore affect grounding-specific mechanism
analyses but cannot inflate resolved counts, since the harness is the only
outcome oracle. Execution-grounding claims are scoped to this validated script
construction.
The FP8 Qwen endpoint is not bit-reproducible even at temperature~0. We repeat
the Qwen headline comparison and RQ4 ablation three times; the Claude Haiku
4.5, DeepSeek-V4-Flash, OpenHands, and TerminalWorld evaluations are single
runs under compute and API-budget constraints, read as per-model,
per-framework, and per-benchmark generality probes.
PMCoder's thresholds and budgets were fixed from pilots before the final
Verified-500 runs, not tuned on the final headline resolved counts.

\subsection{Construct Validity}

SWE-bench Verified resolved status is an operational proxy for issue
resolution. A patch can pass the held-out tests while diverging from maintainer
intent, a known risk in generate-and-validate repair~\cite{qi-patch-plausibility,
smith-overfitting-2015, patch-sim}. We therefore claim improvement on the
official benchmark metric, not human-judged patch quality.
RQ4 isolates the implemented planner and memory components under the fixed
scaffold. It supports plan--memory coupling, but does not separately estimate
the contribution of execution grounding or edit-integrity recovery.

\subsection{External Validity}

The headline evidence comes from SWE-bench Verified-500 and one open-weight 30B
model under a harness-matched baseline. Transfer to other languages and private
codebases remains untested. The API-model,
OpenHands, and TerminalWorld results provide supporting generality evidence, not
a universal SWE-agent ranking.



\section{Related Work}\label{sec:related}

\subsection{Planning and Memory for LLM Agents}

LLM agents are usually given explicit planning or explicit memory, rarely the
two as a coupled design. Planning-centric work decomposes and searches:
plan-and-solve prompting~\cite{wang-plan-and-solve-2023},
hierarchical planner--executor designs~\cite{prasad-adapt-2024, sun-adaplanner-2023},
and, for code, structure-aware search and dependency-graph
planning~\cite{auto-code-rover, codeplan, swe-search};
they maintain progress through an unstructured transcript or a search-specific
control structure, rather than through a memory-coupled plan state.
Memory-centric work expands what the agent can retrieve: reflective and
long-horizon stores~\cite{reflexion, park-generative-agents-2023, packer-memgpt-2023, wang-voyager-2023, xu-amem-2025},
and, for issue resolution, experience memories distilled from past
attempts~\cite{chen-swe-exp-2025, mu-experepair-2025, wong-confucius-2025};
planning there stays implicit or pipeline-fixed. A few designs couple the two,
but outside our setting or in one direction only. Retrieval-augmented planning
and subgoal-chunked working memory are general and non-coding~\cite{kagaya-rap-2024, hu-hiagent-2025}.
Closer to our setting, a recent issue-resolution memory system aligns retrieval
with the plan's sub-task structure but does not let memory drive
replanning~\cite{shen-subtask-memory}, while an experience-bank agent reports a
memory ``synergy'' with test-time sampling rather than planner
coupling~\cite{reasoningbank}.
PMCoder builds on planning and memory work but couples the two: memory-derived trajectory
evidence drives replanning, and verification-phase plan state stays tied to
execution evidence. This makes the design a mechanism study of plan--memory
coupling rather than a leaderboard comparison against systems with different
models, harnesses, and orchestration pipelines.

\subsection{Execution Oracles in Program Repair}

Automated program repair has validated fixes by execution since GenProg accepted
patches that pass the available tests~\cite{genprog},
and the same lineage showed the oracle is weak, as test-passing patches are
often merely plausible~\cite{qi-patch-plausibility, patch-sim}.
Reproduce-then-fix systems turn an issue reproduction into a patch selector that
ranks or filters finished candidates~\cite{libro, agentless, ruan-specrover-2025}.
In all of this the object judged is a completed patch. PMCoder moves the
same kind of behavioral evidence one level earlier, applying it to the agent's
internal plan state during the episode rather than only to a finished patch at
the output.



\section{Conclusion}\label{sec:conclusion}

Our results show that state loss is a consequential failure mode in long
repository-level repair episodes, alongside the base model's reasoning limits.
PMCoder addresses this
directly by coupling a hierarchical phase planner with an episodic memory:
the plan conditions what memory retrieves, and memory-derived signals decide when
to replan. Because such a shared state is only as reliable as the signal that
advances it, a supporting layer grounds verification transitions in
issue-reproduction outcomes instead of the model's self-reported success. On
SWE-bench Verified, this coupling resolves an average of $25$ more issues than a
harness-matched baseline ($+5.0$pp). These results suggest that long-horizon
repair benefits from planning and memory being designed as a coupled state
system, with execution evidence keeping that shared state reliable.

\bibliographystyle{IEEEtran}

\bibliography{references}

@article{agentless,
  title        = {Agentless: Demystifying {LLM}-based Software Engineering Agents},
  author       = {Xia, Chunqiu Steven and Deng, Yinlin and Dunn, Soren and Zhang, Lingming},
  journal      = {Proceedings of the ACM on Software Engineering},
  volume       = {2},
  number       = {FSE},
  pages        = {801--824},
  year         = {2025},
  publisher    = {Association for Computing Machinery},
  doi          = {10.1145/3715754},
  eprint       = {2407.01489},
  archivePrefix = {arXiv},
  primaryClass = {cs.SE},
  url          = {https://arxiv.org/abs/2407.01489},
  note         = {Proceedings of the 33rd ACM SIGSOFT International Symposium on the Foundations of Software Engineering (FSE 2025)}
}

@inproceedings{auto-code-rover,
  author    = {Yuntong Zhang and Haifeng Ruan and Zhiyu Fan and Abhik Roychoudhury},
  title     = {AutoCodeRover: Autonomous Program Improvement},
  booktitle = {Proceedings of the 33rd ACM SIGSOFT International Symposium on Software Testing and Analysis (ISSTA 2024)},
  pages     = {1592--1604},
  year      = {2024},
  publisher = {ACM},
  address   = {Vienna, Austria},
  doi       = {10.1145/3650212.3680384},
  eprint    = {2404.05427},
  archivePrefix = {arXiv},
  url       = {https://arxiv.org/abs/2404.05427}
}

@inproceedings{carbonell-goldstein-mmr,
  author    = {Carbonell, Jaime and Goldstein, Jade},
  title     = {The Use of {MMR}, Diversity-Based Reranking for Reordering Documents and Producing Summaries},
  booktitle = {Proceedings of the 21st Annual International ACM SIGIR Conference on Research and Development in Information Retrieval (SIGIR '98)},
  pages     = {335--336},
  year      = {1998},
  publisher = {ACM},
  address   = {New York, NY, USA},
  doi       = {10.1145/290941.291025},
  url       = {https://doi.org/10.1145/290941.291025}
}

@article{clark-chalmers-extended-mind,
  author  = {Clark, Andy and Chalmers, David J.},
  title   = {The Extended Mind},
  journal = {Analysis},
  volume  = {58},
  number  = {1},
  pages   = {7--19},
  year    = {1998},
  doi     = {10.1093/analys/58.1.7},
  url     = {https://doi.org/10.1093/analys/58.1.7}
}

@article{genprog,
  author       = {Le Goues, Claire and Nguyen, ThanhVu and Forrest, Stephanie and Weimer, Westley},
  title        = {{GenProg}: A Generic Method for Automatic Software Repair},
  journal      = {IEEE Transactions on Software Engineering},
  volume       = {38},
  number       = {1},
  pages        = {54--72},
  year         = {2012},
  doi          = {10.1109/TSE.2011.104},
  url          = {https://doi.org/10.1109/TSE.2011.104}
}

@inproceedings{gou-critic-2024,
  title     = {{CRITIC}: Large Language Models Can Self-Correct with Tool-Interactive Critiquing},
  author    = {Gou, Zhibin and Shao, Zhihong and Gong, Yeyun and Shen, Yelong and Yang, Yujiu and Duan, Nan and Chen, Weizhu},
  booktitle = {The Twelfth International Conference on Learning Representations (ICLR)},
  year      = {2024},
  eprint    = {2305.11738},
  archivePrefix = {arXiv},
  primaryClass = {cs.CL},
  url       = {https://arxiv.org/abs/2305.11738}
}

@inproceedings{huang-self-correct-2024,
  title     = {Large Language Models Cannot Self-Correct Reasoning Yet},
  author    = {Huang, Jie and Chen, Xinyun and Mishra, Swaroop and Zheng, Huaixiu Steven and Yu, Adams Wei and Song, Xinying and Zhou, Denny},
  booktitle = {The Twelfth International Conference on Learning Representations (ICLR)},
  year      = {2024},
  url       = {https://openreview.net/forum?id=IkmD3fKBPQ},
  eprint    = {2310.01798},
  archivePrefix = {arXiv},
  primaryClass = {cs.CL}
}

@inproceedings{kwon-vllm,
  author    = {Woosuk Kwon and Zhuohan Li and Siyuan Zhuang and Ying Sheng and Lianmin Zheng and Cody Hao Yu and Joseph E. Gonzalez and Hao Zhang and Ion Stoica},
  title     = {Efficient Memory Management for Large Language Model Serving with {PagedAttention}},
  booktitle = {Proceedings of the 29th Symposium on Operating Systems Principles (SOSP)},
  pages     = {611--626},
  year      = {2023},
  publisher = {ACM},
  doi       = {10.1145/3600006.3613165},
  eprint    = {2309.06180},
  archivePrefix = {arXiv},
  url       = {https://arxiv.org/abs/2309.06180}
}

@inproceedings{libro,
  author    = {Kang, Sungmin and Yoon, Juyeon and Yoo, Shin},
  title     = {Large Language Models are Few-shot Testers: Exploring {LLM}-based General Bug Reproduction},
  booktitle = {Proceedings of the 45th International Conference on Software Engineering (ICSE)},
  series    = {ICSE '23},
  publisher = {IEEE Press},
  year      = {2023},
  pages     = {2312--2323},
  doi       = {10.1109/ICSE48619.2023.00194},
  eprint    = {2209.11515},
  archivePrefix = {arXiv},
  primaryClass  = {cs.SE},
  url       = {https://arxiv.org/abs/2209.11515}
}

@inproceedings{openhands,
  title     = {{OpenHands}: An Open Platform for {AI} Software Developers as Generalist Agents},
  author    = {Wang, Xingyao and Li, Boxuan and Song, Yufan and Xu, Frank F. and Tang, Xiangru and Zhuge, Mingchen and Pan, Jiayi and Song, Yueqi and Li, Bowen and Singh, Jaskirat and Tran, Hoang H. and Li, Fuqiang and Ma, Ren and Zheng, Mingzhang and Qian, Bill and Shao, Yanjun and Muennighoff, Niklas and Zhang, Yizhe and Hui, Binyuan and Lin, Junyang and Brennan, Robert and Peng, Hao and Ji, Heng and Neubig, Graham},
  booktitle = {International Conference on Learning Representations (ICLR)},
  year      = {2025},
  eprint    = {2407.16741},
  archivePrefix = {arXiv},
  primaryClass  = {cs.SE},
  url       = {https://openreview.net/forum?id=OJd3ayDDoF}
}

@article{packer-memgpt-2023,
  title         = {{MemGPT}: Towards {LLMs} as Operating Systems},
  author        = {Packer, Charles and Wooders, Sarah and Lin, Kevin and Fang, Vivian and Patil, Shishir G. and Stoica, Ion and Gonzalez, Joseph E.},
  year          = {2023},
  journal       = {arXiv preprint arXiv:2310.08560},
  eprint        = {2310.08560},
  archivePrefix = {arXiv},
  primaryClass  = {cs.AI},
  url           = {https://arxiv.org/abs/2310.08560}
}

@inproceedings{park-generative-agents-2023,
  author    = {Park, Joon Sung and O'Brien, Joseph and Cai, Carrie Jun and Morris, Meredith Ringel and Liang, Percy and Bernstein, Michael S.}, 
  title     = {Generative Agents: Interactive Simulacra of Human Behavior},
  year      = {2023},
  isbn      = {9798400701320},
  publisher = {Association for Computing Machinery},
  address   = {New York, NY, USA},
  booktitle = {Proceedings of the 36th Annual ACM Symposium on User Interface Software and Technology},
  series    = {UIST '23},
  pages     = {1--22},
  doi       = {10.1145/3586183.3606763},
  url       = {https://doi.org/10.1145/3586183.3606763},
}

@inproceedings{patch-sim,
  author    = {Yingfei Xiong and Xinyuan Liu and Muhan Zeng and Lu Zhang and Gang Huang},
  title     = {Identifying Patch Correctness in Test-Based Program Repair},
  booktitle = {Proceedings of the 40th International Conference on Software Engineering (ICSE)},
  pages     = {789--799},
  year      = {2018},
  publisher = {ACM},
  address   = {Gothenburg, Sweden},
  doi       = {10.1145/3180155.3180182},
  url       = {https://doi.org/10.1145/3180155.3180182}
}

@inproceedings{qi-patch-plausibility,
  author    = {Qi, Zichao and Long, Fan and Achour, Sara and Rinard, Martin},
  title     = {An Analysis of Patch Plausibility and Correctness for Generate-and-Validate Patch Generation Systems},
  booktitle = {Proceedings of the 2015 International Symposium on Software Testing and Analysis (ISSTA 2015)},
  pages     = {24--36},
  year      = {2015},
  publisher = {ACM},
  address   = {Baltimore, MD, USA},
  doi       = {10.1145/2771783.2771791},
  url       = {https://doi.org/10.1145/2771783.2771791}
}

@misc{qwen3technicalreport,
  title={Qwen3 Technical Report},
  author={Qwen Team},
  year={2025},
  eprint={2505.09388},
  archivePrefix={arXiv},
  primaryClass={cs.CL},
  url={https://arxiv.org/abs/2505.09388},
}

@inproceedings{reflexion,
  title     = {Reflexion: Language Agents with Verbal Reinforcement Learning},
  author    = {Shinn, Noah and Cassano, Federico and Berman, Edward and Gopinath, Ashwin and Narasimhan, Karthik and Yao, Shunyu},
  booktitle = {Advances in Neural Information Processing Systems 36 (NeurIPS 2023)},
  year      = {2023},
  eprint    = {2303.11366},
  archivePrefix = {arXiv},
  primaryClass = {cs.AI},
  doi       = {10.48550/arXiv.2303.11366},
  url       = {https://arxiv.org/abs/2303.11366}
}

@inproceedings{ruan-specrover-2025,
  title        = {SpecRover: Code Intent Extraction via LLMs},
  author       = {Ruan, Haifeng and Zhang, Yuntong and Roychoudhury, Abhik},
  booktitle    = {2025 IEEE/ACM 47th International Conference on Software Engineering (ICSE)},
  pages        = {963--974},
  year         = {2025},
  doi          = {10.1109/ICSE55347.2025.00080},
  eprint       = {2408.02232},
  archivePrefix = {arXiv},
  url          = {https://arxiv.org/abs/2408.02232}
}

@inproceedings{smith-overfitting-2015,
  author    = {Smith, Edward K. and Barr, Earl T. and Le Goues, Claire and Brun, Yuriy},
  title     = {Is the Cure Worse Than the Disease? Overfitting in Automated Program Repair},
  booktitle = {Proceedings of the 2015 10th Joint Meeting on Foundations of Software Engineering (ESEC/FSE 2015)},
  year      = {2015},
  pages     = {532--543},
  publisher = {ACM},
  address   = {Bergamo, Italy},
  doi       = {10.1145/2786805.2786825},
  url       = {https://doi.org/10.1145/2786805.2786825}
}

@inproceedings{yang2024sweagent,
  title={{SWE}-agent: Agent-Computer Interfaces Enable Automated Software Engineering},
  author={John Yang and Carlos E Jimenez and Alexander Wettig and Kilian Lieret and Shunyu Yao and Karthik R Narasimhan and Ofir Press},
  booktitle={The Thirty-eighth Annual Conference on Neural Information Processing Systems},
  year={2024},
  url={https://arxiv.org/abs/2405.15793}
}

@inproceedings{swe-bench,
  title     = {{SWE}-bench: Can Language Models Resolve Real-world {Github} Issues?},
  author    = {Carlos E Jimenez and John Yang and Alexander Wettig and Shunyu Yao and Kexin Pei and Ofir Press and Karthik R Narasimhan},
  booktitle = {The Twelfth International Conference on Learning Representations (ICLR)},
  year      = {2024},
  eprint    = {2310.06770},
  archivePrefix = {arXiv},
  primaryClass = {cs.CL},
  url       = {https://openreview.net/forum?id=VTF8yNQM66}
}

@inproceedings{wang-plan-and-solve-2023,
    title = {Plan-and-Solve Prompting: Improving Zero-Shot Chain-of-Thought Reasoning by Large Language Models},
    author = {Wang, Lei and Xu, Wanyu and Lan, Yihuai and Hu, Zhiqiang and Lan, Yunshi and Lee, Roy Ka-Wei and Lim, Ee-Peng},
    booktitle = {Proceedings of the 61st Annual Meeting of the Association for Computational Linguistics (Volume 1: Long Papers)},
    month = jul,
    year = {2023},
    address = {Toronto, Canada},
    publisher = {Association for Computational Linguistics},
    pages = {2609--2634},
    doi = {10.18653/v1/2023.acl-long.147},
    url = {https://aclanthology.org/2023.acl-long.147/},
    eprint = {2305.04091},
    archivePrefix = {arXiv}
}

@article{wang-voyager-2023,
  title        = {Voyager: An Open-Ended Embodied Agent with Large Language Models},
  author       = {Wang, Guanzhi and Xie, Yuqi and Jiang, Yunfan and Mandlekar, Ajay and Xiao, Chaowei and Zhu, Yuke and Fan, Linxi and Anandkumar, Anima},
  journal      = {Transactions on Machine Learning Research (TMLR)},
  year         = {2024},
  eprint       = {2305.16291},
  archivePrefix = {arXiv},
  primaryClass = {cs.AI},
  url          = {https://arxiv.org/abs/2305.16291}
}

@article{zeller-hildebrandt-delta-debugging,
  author    = {Zeller, Andreas and Hildebrandt, Ralf},
  title     = {Simplifying and Isolating Failure-Inducing Input},
  journal   = {IEEE Transactions on Software Engineering},
  volume    = {28},
  number    = {2},
  pages     = {183--200},
  year      = {2002},
  publisher = {IEEE},
  doi       = {10.1109/32.988498},
  url       = {https://doi.org/10.1109/32.988498}
}

@misc{mu-experepair-2025,
  title        = {{EXPEREPAIR}: Dual-Memory Enhanced LLM-based Repository-Level Program Repair},
  author       = {Mu, Fangwen and Wang, Junjie and Shi, Lin and Wang, Song and Li, Shoubin and Wang, Qing},
  year         = {2026},
  eprint       = {2506.10484},
  archivePrefix= {arXiv},
  primaryClass = {cs.SE},
  note         = {Accepted by FSE 2026},
  doi          = {10.48550/arXiv.2506.10484},
  url          = {https://arxiv.org/abs/2506.10484}
}

@misc{chen-swe-exp-2025,
  title        = {SWE-Exp: Experience-Driven Software Issue Resolution},
  author       = {Chen, Silin and Lin, Shaoxin and Shi, Yuling and Lian, Heng and Gu, Xiaodong and Yun, Longfei and Chen, Dong and Cao, Lin and Liu, Jiyang and Xia, Nu and Wang, Qianxiang},
  year         = {2025},
  eprint       = {2507.23361},
  archivePrefix = {arXiv},
  primaryClass = {cs.SE},
  doi          = {10.48550/arXiv.2507.23361},
  url          = {https://arxiv.org/abs/2507.23361}
}

@inproceedings{hu-hiagent-2025,
    title = "{H}i{A}gent: Hierarchical Working Memory Management for Solving Long-Horizon Agent Tasks with Large Language Model",
    author = "Hu, Mengkang and Chen, Tianxing and Chen, Qiguang and Mu, Yao and Shao, Wenqi and Luo, Ping",
    editor = "Che, Wanxiang and Nabende, Joyce and Shutova, Ekaterina and Pilehvar, Mohammad Taher",
    booktitle = "Proceedings of the 63rd Annual Meeting of the Association for Computational Linguistics (Volume 1: Long Papers)",
    month = jul,
    year = "2025",
    address = "Vienna, Austria",
    publisher = "Association for Computational Linguistics",
    url = "https://aclanthology.org/2025.acl-long.1575/",
    doi = "10.18653/v1/2025.acl-long.1575",
    pages = "32779--32798",
    ISBN = "979-8-89176-251-0"
}

@misc{kagaya-rap-2024,
  title        = {{RAP}: Retrieval-Augmented Planning with Contextual Memory for Multimodal {LLM} Agents},
  author       = {Kagaya, Tomoyuki and Yuan, Thong Jing and Lou, Yuxuan and Karlekar, Jayashree and Pranata, Sugiri and Kinose, Akira and Oguri, Koki and Wick, Felix and You, Yang},
  year         = {2024},
  eprint       = {2402.03610},
  archivePrefix = {arXiv},
  primaryClass = {cs.LG},
  url          = {https://arxiv.org/abs/2402.03610}
}

@misc{wong-confucius-2025,
  title        = {Confucius Code Agent: Scalable Agent Scaffolding for Real-World Codebases},
  author       = {Wong, Sherman and Qi, Zhenting and Wang, Zhaodong and Hu, Nathan and Lin, Samuel and Ge, Jun and Gao, Erwin and Chen, Wenlin and Du, Yilun and Yu, Minlan and Zhang, Ying},
  year         = {2025},
  eprint       = {2512.10398},
  archivePrefix = {arXiv},
  primaryClass = {cs.CL},
  doi          = {10.48550/arXiv.2512.10398},
  url          = {https://arxiv.org/abs/2512.10398}
}

@inproceedings{xu-amem-2025,
  title     = {A-{MEM}: Agentic Memory for {LLM} Agents},
  author    = {Xu, Wujiang and Liang, Zujie and Mei, Kai and Gao, Hang and Tan, Juntao and Zhang, Yongfeng},
  booktitle = {Advances in Neural Information Processing Systems (NeurIPS)},
  year      = {2025},
  eprint    = {2502.12110},
  archivePrefix = {arXiv},
  primaryClass  = {cs.CL},
  doi       = {10.48550/arXiv.2502.12110},
  url       = {https://arxiv.org/abs/2502.12110}
}

@article{denison-reward-tampering-2024,
  title         = {Sycophancy to Subterfuge: Investigating Reward-Tampering in Large Language Models},
  author        = {Denison, Carson and MacDiarmid, Monte and Barez, Fazl and Duvenaud, David and Kravec, Shauna and Marks, Samuel and Schiefer, Nicholas and Soklaski, Ryan and Tamkin, Alex and Kaplan, Jared and Shlegeris, Buck and Bowman, Samuel R. and Perez, Ethan and Hubinger, Evan},
  year          = {2024},
  journal       = {arXiv preprint arXiv:2406.10162},
  eprint        = {2406.10162},
  archivePrefix = {arXiv},
  primaryClass  = {cs.AI},
  url           = {https://arxiv.org/abs/2406.10162}
}

@inproceedings{pan-reward-misspec-2022,
  title     = {The Effects of Reward Misspecification: Mapping and Mitigating Misaligned Models},
  author    = {Pan, Alexander and Bhatia, Kush and Steinhardt, Jacob},
  booktitle = {The Tenth International Conference on Learning Representations (ICLR)},
  year      = {2022},
  url       = {https://openreview.net/forum?id=JYtwGwIL7ye},
  eprint    = {2201.03544},
  archivePrefix = {arXiv},
  primaryClass  = {cs.LG}
}

@inproceedings{skalse-reward-hacking-2022,
  title     = {Defining and Characterizing Reward Hacking},
  author    = {Skalse, Joar and Howe, Nikolaus H. R. and Krasheninnikov, Dmitrii and Krueger, David},
  booktitle = {Advances in Neural Information Processing Systems 35 (NeurIPS 2022)},
  year      = {2022},
  eprint    = {2209.13085},
  archivePrefix = {arXiv},
  primaryClass  = {cs.LG},
  url       = {https://arxiv.org/abs/2209.13085}
}

@misc{sun-adaplanner-2023,
  title={AdaPlanner: Adaptive Planning from Feedback with Language Models}, 
  author={Haotian Sun and Yuchen Zhuang and Lingkai Kong and Bo Dai and Chao Zhang},
  year={2023},
  eprint={2305.16653},
  archivePrefix={arXiv},
  primaryClass={cs.CL},
  url={https://arxiv.org/abs/2305.16653}, 
}

@inproceedings{prasad-adapt-2024,
    title = "{AD}a{PT}: As-Needed Decomposition and Planning with Language Models",
    author = "Prasad, Archiki  and
      Koller, Alexander  and
      Hartmann, Mareike  and
      Clark, Peter  and
      Sabharwal, Ashish  and
      Bansal, Mohit  and
      Khot, Tushar",
    editor = "Duh, Kevin  and
      Gomez, Helena  and
      Bethard, Steven",
    booktitle = "Findings of the Association for Computational Linguistics: NAACL 2024",
    month = jun,
    year = "2024",
    address = "Mexico City, Mexico",
    publisher = "Association for Computational Linguistics",
    url = "https://aclanthology.org/2024.findings-naacl.264/",
    doi = "10.18653/v1/2024.findings-naacl.264",
    pages = "4226--4252"
}

@misc{swe-bench-verified,
  author       = {{OpenAI}},
  title        = {Introducing {SWE}-bench {V}erified},
  year         = {2024},
  howpublished = {OpenAI},
  url          = {https://openai.com/index/introducing-swe-bench-verified/}
}

@article{sumers-coala,
  title         = {Cognitive Architectures for Language Agents},
  author        = {Theodore R. Sumers and Shunyu Yao and Karthik Narasimhan and Thomas L. Griffiths},
  journal       = {Transactions on Machine Learning Research (TMLR)},
  year          = {2024},
  eprint        = {2309.02427},
  archivePrefix = {arXiv},
  primaryClass  = {cs.AI},
  url           = {https://arxiv.org/abs/2309.02427}
}

@misc{wang-se-agent-survey,
  title         = {Agents in Software Engineering: Survey, Landscape, and Vision},
  author        = {Yanlin Wang and Wanjun Zhong and Yanxian Huang and Ensheng Shi and Min Yang and Jiachi Chen and Hui Li and Yuchi Ma and Qianxiang Wang and Zibin Zheng},
  year          = {2024},
  eprint        = {2409.09030},
  archivePrefix = {arXiv},
  primaryClass  = {cs.SE},
  url           = {https://arxiv.org/abs/2409.09030}
}

@article{codeplan,
author = {Bairi, Ramakrishna and Sonwane, Atharv and Kanade, Aditya and C., Vageesh D. and Iyer, Arun and Parthasarathy, Suresh and Rajamani, Sriram and Ashok, B. and Shet, Shashank},
title = {CodePlan: Repository-Level Coding using LLMs and Planning},
year = {2024},
issue_date = {July 2024},
publisher = {Association for Computing Machinery},
address = {New York, NY, USA},
volume = {1},
number = {FSE},
url = {https://doi.org/10.1145/3643757},
doi = {10.1145/3643757},
journal = {Proceedings of the ACM on Software Engineering},
month = jul,
articleno = {31},
numpages = {24}
}

@inproceedings{swe-search,
  title     = {{SWE}-Search: Enhancing Software Agents with Monte Carlo Tree Search and Iterative Refinement},
  author    = {Antonis Antoniades and Albert {\"O}rwall and Kexun Zhang and Yuxi Xie and Anirudh Goyal and William Yang Wang},
  booktitle = {The Thirteenth International Conference on Learning Representations (ICLR)},
  year      = {2025},
  url       = {https://openreview.net/forum?id=G7sIFXugTX}
}

@inproceedings{reasoningbank,
  title         = {{ReasoningBank}: Scaling Agent Self-Evolving with Reasoning Memory},
  author        = {Siru Ouyang and Jun Yan and I-Hung Hsu and Yanfei Chen and Ke Jiang and Zifeng Wang and Rujun Han and Long T. Le and Samira Daruki and Xiangru Tang and Vishy Tirumalashetty and George Lee and Mahsan Rofouei and Hangfei Lin and Jiawei Han and Chen-Yu Lee and Tomas Pfister},
  booktitle     = {International Conference on Learning Representations (ICLR)},
  year          = {2026},
  eprint        = {2509.25140},
  archivePrefix = {arXiv},
  primaryClass  = {cs.CL},
  doi           = {10.48550/arXiv.2509.25140},
  url           = {https://arxiv.org/abs/2509.25140}
}

@misc{shen-subtask-memory,
  title         = {Structurally Aligned Subtask-Level Memory for Software Engineering Agents},
  author        = {Kangning Shen and Jingyuan Zhang and Chenxi Sun and Wencong Zeng and Yang Yue},
  year          = {2026},
  eprint        = {2602.21611},
  archivePrefix = {arXiv},
  primaryClass  = {cs.SE},
  doi           = {10.48550/arXiv.2602.21611},
  url           = {https://arxiv.org/abs/2602.21611}
}

@misc{kaddour-overconfidence-2026,
  title         = {Agentic Uncertainty Reveals Agentic Overconfidence},
  author        = {Kaddour, Jean and Patel, Srijan and Dovonon, Gbètondji and Richter, Leo and Minervini, Pasquale and Kusner, Matt J.},
  year          = {2026},
  eprint        = {2602.06948},
  archivePrefix = {arXiv},
  primaryClass  = {cs.AI},
  url           = {https://arxiv.org/abs/2602.06948}
}

@misc{terminalworld,
  title         = {{TerminalWorld}: Benchmarking Agents on Real-World Terminal Tasks},
  author        = {Chu, Zhaoyang and Hu, Jiarui and Jiang, Xingyu and Zou, Pengyu and Li, Han and Peng, Chao and O'Hearn, Peter and Barr, Earl T. and Harman, Mark and Sarro, Federica and Ye, He},
  year          = {2026},
  eprint        = {2605.22535},
  archivePrefix = {arXiv},
  primaryClass  = {cs.SE},
  doi           = {10.48550/arXiv.2605.22535},
  url           = {https://arxiv.org/abs/2605.22535}
}

@book{hutchins-distributed-cognition-1995,
  author    = {Hutchins, Edwin},
  title     = {Cognition in the {W}ild},
  publisher = {MIT Press},
  address   = {Cambridge, MA},
  year      = {1995}
}

@article{risko-gilbert-offloading-2016,
  author  = {Risko, Evan F. and Gilbert, Sam J.},
  title   = {Cognitive Offloading},
  journal = {Trends in Cognitive Sciences},
  volume  = {20},
  number  = {9},
  pages   = {676--688},
  year    = {2016},
  doi     = {10.1016/j.tics.2016.07.002}
}

\end{document}